\documentclass[12pt,twoside]{article}
\usepackage{fleqn,espcrc1}

\usepackage{graphicx}
\usepackage{epsfig}
\usepackage[figuresright]{rotating}

\newcommand{\AmS}{{\protect\the\textfont2
  A\kern-.1667em\lower.5ex\hbox{M}\kern-.125emS}}

\title{Quark model description of the $NN^*$(1440) potential}

\author{B. Juli\'a$^{\rm a}$, F. Fern\'andez$^{\rm a}$,
A. Valcarce\address{Grupo de F\'\i sica Nuclear, Universidad
de Salamanca,\\
 E-37008 Salamanca, Spain \\ },
        and
        P. Gonz\'alez\address{Dpto. de F\'\i sica Te\'orica-IFIC, Universidad
de Valencia-CSIC,\\
       E-46100 Burjassot, Valencia, Spain }}%
       
\begin{document}

% typeset front matter
\maketitle

\begin{abstract}
We derive a $NN^*$(1440) potential from a non-relativistic
quark-quark interaction and a chiral quark cluster model for 
the baryons.
By making use of the Born-Oppenheimer approximation we
examine the most important features of this interaction
in comparison to those obtained from meson-exchange
models.

\end{abstract}

\section{INTRODUCTION}
Baryonic resonances play a major role in the 
understanding of reactions that take place in nucleons
and nuclei in the so-called intermediate energy regime \cite{Oset82}.
In particular, the low-lying nucleonic resonances $\Delta$(1232)
and $N^*$(1440), can be now analyzed in more detail due to 
the development of specific experimental programs in TJNAF, Uppsala...

In this context the transition, $NN\rightarrow NR$ \ ($R$ : resonance), and
direct $NR\rightarrow NR$ and $RR\rightarrow RR$ interactions should be
understood. Usually these interactions have been written as straightforward
extensions of some pieces of the $NN\rightarrow NN$ potential with the
modification of the values of the coupling constants, extracted from their
decay widths. Though this procedure can be appropriate for the very
long-range part of the interaction, it is under suspicion at least for the
short-range part for which the detailed structure of the baryons may
determine to some extent the form of the interaction. This turns out to be
the case for the $NN\rightarrow N\Delta $ and $N\Delta \rightarrow N\Delta $
potentials previously analyzed elsewhere \cite{Valc95}. It seems therefore
convenient to proceed to a derivation of these potentials based on the more
elementary quark-quark interaction. 

This is the purpose of this talk:
starting from a quark-quark non-relativistic interaction, we implement the
baryon structure through technically simple gaussian wave
functions and we calculate the potential at the baryonic level in the static
Born-Oppenheimer approach. The $N^{\ast }(1440)$, the Roper resonance, 
considered as a radial excitation of the nucleon, is
taken as a stable particle. For dynamical applications its width should be
implemented through the coupling to the continuum.

 We center our attention
in the $NN^{\ast }\rightarrow NN^{\ast }$ potential where a complete
parallelism with the $NN\rightarrow NN$ case can be easily established.
Notice that the quark-quark interaction parameters are fixed (from the $%
NN\rightarrow NN$ case) and are kept independent of the baryons involved in
the interaction. This eliminates the bias introduced in models at the
baryonic level by a different choice of effective parameters according to
the baryon-baryon interaction considered (this effectiveness of the
parameters may hide distinct physical effects).

\section{THE $NN^*$(1440) WAVE FUNCTION}

The wave function of a two-baryon system, $B_1$ and $B_2$,
 with a definite symmetry under the exchange
of the baryon quantum numbers is written as \cite{QP}:

\begin{eqnarray}
\Psi_{B_1 B_2}^{ST}({\vec R}) & = & {\frac{{\cal A} }{\sqrt{1 + \delta_{B_1
B_2}}}} \sqrt{\frac{1 }{2}} \Biggr\{ \left[ B_1 \left( 123;{-{\frac{{\vec R} 
}{2}}} \right) B_2 \left( 456; {\frac{{\vec R} }{2}} \right) \right]_{ST} 
\nonumber \\
& + & (-1)^{f} \, \Biggr\{ \left[ B_2 \left( 123;{-{\frac{{\vec R} }{2}}}
\right) B_1 \left( 456; {\frac{{\vec R} }{2}} \right) \right]_{ST} \Biggr \} %
\, ,  \label{Gor}
\end{eqnarray}

\noindent being ${\cal A}$ the six-quark antisymmetrizer given by:
\begin{equation}
{\cal A} = (1-\sum_{i=1}^3\sum_{j=4}^6 P_{ij})(1-{\cal P}) \, ,
\end{equation}

\noindent where ${\cal P}$ exchanges the three quarks between the
two clusters and $P_{ij}$ exchanges
quarks $i$ and $j$.

If one projects on a state of definite orbital angular momentum $L$, due to
the $(1-{\cal P})$ operator in the antisymmetrizer the wave function $%
\Psi_{B_1 B_2}^{ST}({\vec R})$ vanishes unless:

\begin{equation}
L+S_{1}+S_{2}-S+T_{1}+T_{2}-T+f = {\rm odd} \, .
\end{equation}

\noindent Since $S_{1}={\frac{1 }{2}} =S_{2}$, $T_{1}={\frac{1 }{2}} =T_{2}$%
, this fixes the relative phase between the two components of the wave
function at Eq. (\ref{Gor}) to be:

\begin{equation}
f=S+T-L+{\rm odd}\,.
\end{equation}

It is important to realize that for the $NN$ system $f$ is necessarily even
in order to prevent the vanishing of the wave function. No such restriction
exists for $NN^{\ast }$. Therefore, there are $NN^{\ast }$ channels ($f$
odd) with no counterpart in the $NN$ case.

We will assume the three-quark
wave function for the quark clusters at a position $\vec{R}$ 
to be given by

\begin{equation}
|N\rangle = |[3](0s)^3\rangle\, ,
\end{equation}

\begin{equation}
|N^*\rangle = \sqrt{2\over3}|[3](0s)^2(1s)\rangle
-\sqrt{1\over3} |[3](0s)(op)^2\rangle \, ,
\end{equation}

\noindent 
explicitly, 
\begin{equation}
N({\vec r}_1,{\vec r}_2,{\vec r}_3;\vec{R}) = \prod_{n=1}^{3} \left({\frac{1 
}{\pi b^{2}}} \right)^{3/4} e^{-{\frac{ ( \vec{r}_{n}- \vec{R})^2 }{2 b^{2} }}
} \otimes [3]_{ST} \otimes [1^{3}]_{c} \, ,
\end{equation}

\noindent 
and
\begin{equation}
N^{*}({\vec r}_1,{\vec r}_2,{\vec r}_3;\vec{R}) = 
( \sqrt{2\over3} \phi_1- \sqrt{1\over3}\phi_2 )
\otimes[3]_{ST}\otimes[1^3]_C \, ,
\end{equation}

\noindent 
being
\begin{equation}
\phi_1 =  {\sqrt{2} \over 3}
\left( {\frac{1 }{\pi b^{2} }} \right)^{9/4}
\sum_{k=1}^3 \left[ {\frac{3 }{2}} - {\frac{ ( \vec{r}_k%
- \vec{R})^2 }{b^{2} }}\right] \prod_{i=1}^3 e^{-{\frac{ ( \vec{r}_{i}- \vec{R})^2 }{2
b^{2} }} } ,
\end{equation}

\noindent
and
\begin{equation}
\phi_2 =  -{2 \over 3}
\left( 1\over \pi^{9\over4} b^{13\over2}  \right)  
\sum_{j<k=1}^3 (\vec{r}_j-\vec{R})\cdot (\vec{r}_k-\vec{R})
 \prod_{i=1}^3 e^{-{\frac{ ( \vec{r}_{i}- \vec{R})^2 }{2
b^{2} }} } ,
\end{equation}
\noindent where $[3]_{ST}$ and $[1^{3}]_{c}$ stand for the spin-isospin and
color part, respectively.
The validity of the harmonic oscillator wave functions to calculate 
the two-baryon interaction has been discussed in 
ref. \cite{PLB}.

The quark-quark potential we use can be written in terms of the 
interquark distance $\vec{r}_{ij}$ as:
\begin{equation}
V_{qq}(\vec{r}_{ij})=V_{CON}(\vec{r}_{ij})+V_{OGE}(\vec{r}_{ij})+V_{OPE}(%
\vec{r}_{ij})+V_{OSE}(\vec{r}_{ij})\,,
\ref{POT}
\end{equation}
where $V_{CON}$ stands for the confining potential, and
$V_{OGE}$, $V_{OPE}$, and $V_{OSE}$
for one-gluon, one-pion and one-sigma exchange potentials,
respectively. The expression 
of these potentials has been very much detailed elsewhere \cite{JPG}.

The baryon-baryon potential is obtained as the expectation value 
of the energy of the six-quark system minus the self-energy of the 
two clusters.
The presence of the antisymmetrization in the two-baryon 
wave function has also an important dynamical effect, the 
baryon-baryon potential contains quark-exchange contributions 
where the interaction takes place between two baryons that 
exchange a quark.

\section{RESULTS}
\begin{figure}[htb]
\centering
\scalebox{0.70}{\includegraphics{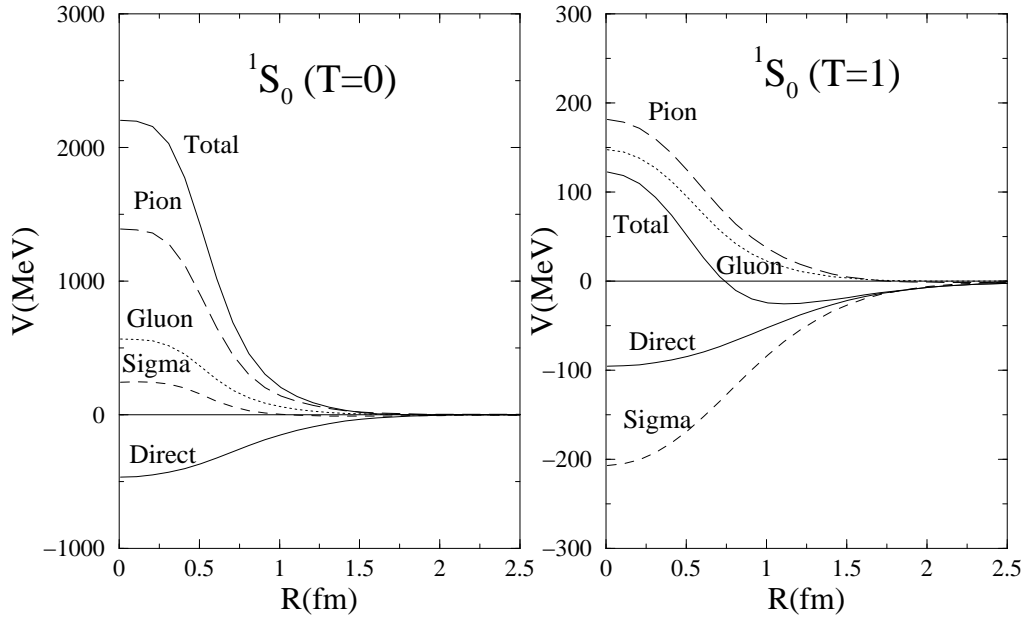}}
\vspace*{-1.00cm}
\caption{$^1S_{0} (T=0$)
and $^1S_0 (T=1)$ $NN^*$ potentials. }
\label{Fig1}
\vspace*{-0.50cm}
\end{figure}

In Figure \ref{Fig1} we show the results for the $NN^*$ potential in terms 
of the interbaryon distance R for two channels: 
 $^1S_{0} (T=0)$, which is forbidden in the $NN$ system, and the
$^1S_{0} (T=1)$, which is allowed in the $NN$ system.
In this last case, the result is quite close to the 
corresponding channel in the $NN$ system, a consequence of the 
near to identity similarity of $N$ and $N^*$. As can
be seen, the behavior in the two previous channels
 is completely different such that it 
could not be obtained by a simple rescaling of the vertex
coupling constants from one case to the other.
In order to emphasize the effects of quark 
antisymmetrization, 
we have compared to a
direct potential without quark-exchange contributions.
We have also separated
the contribution of the different terms of 
the quark-quark potential in Eq. \label{POT} 

As general features of the results we may remark
that the OPE interaction determines the very long range
behavior (R$>$4 fm), the OPE altogether with the OSE are
responsible for the long-range par (1.5 fm $<$ R $<$ 4 fm),
 and OPE, OSE and OGE added to quark-exchange determine 
the attractive or repulsive character of the interaction at 
the intermediate- and short-range.

Certainly data on $NN^* \rightarrow NN^*$ phase shifts can be only obtained
indirectly and no direct experimental test of our results
can actually be performed.
Nonetheless, our results should help
to a better understanding of baryonic processes at a microscopic level and
serve as a guide when dealing with reactions where some indicative
predictions are needed in theoretical as well as in experimental studies.
The elastic $\pi d$ scattering above the Roper threshold as well as the
breakup of the deuteron into $NN^*$ channels, although not available 
for the moment, should serve as a test of the results we have derived.

A transition potential $NN\rightarrow NN^*$ can also be
derived within the same framework. Althought this transition
does not show forbidden channels, the quantum numbers
are fixed by the $NN$ system,
the quark model provides a parameter-free prediction.
This potential can be tested in several reactions \cite{SAT}.
We have determined  the $NN^*$(1440)  probability
on the deuteron by means of a multichannel calculation 
including:
${}^3\!S_1^{NN}$, ${}^3\!D_1^{NN}$, ${}^3\!S_1^{\Delta\Delta}$, 
${}^3\!D_1^{\Delta\Delta}$, ${}^7\!D_1^{\Delta\Delta}$, ${}^7\!G_1^{\Delta\Delta}$
,${}^3\!S_1^{NN^*(1440)}$, and ${}^3\!D_1^{NN^*(1440)} $,
 finding for the Roper components
0.003$\%$ and 0.024$\%$, respectively, much lower than the 
$\Delta\Delta$ ones ($\approx$ 0.25$\%$).

\section{ACKNOWLEDGMENTS}
We thank to D. R. Entem for the calculations on the deuteron.

\end{document}